\documentclass[prd,preprint,nofootinbib]{revtex4}

\usepackage{graphicx}
\usepackage{amssymb}
\usepackage{amsmath}

\preprint{IPMU-10-0102}

\begin{document}

%%%%%%% added by Fumi %%%%%%%%%%
% FROM HERE
\newcommand{\beq}{\begin{equation}}   
\newcommand{\eeq}{\end{equation}}
\newcommand{\bea}{\begin{eqnarray}}   
\newcommand{\eea}{\end{eqnarray}}
\newcommand{\bear}{\begin{array}}  
\newcommand {\eear}{\end{array}}
\newcommand{\bef}{\begin{figure}}  
\newcommand {\eef}{\end{figure}}
\newcommand{\bec}{\begin{center}}  
\newcommand {\eec}{\end{center}}
\newcommand{\non}{\nonumber}  
\newcommand {\eqn}[1]{\beq {#1}\eeq}
\newcommand{\la}{\left\langle}  
\newcommand{\ra}{\right\rangle}
\newcommand{\ds}{\displaystyle}
\def\SEC#1{Sec.~\ref{#1}}
\def\FIG#1{Fig.~\ref{#1}}
\def\EQ#1{Eq.~(\ref{#1})}
\def\EQS#1{Eqs.~(\ref{#1})}
\def\GEV#1{10^{#1}{\rm\,GeV}}
\def\MEV#1{10^{#1}{\rm\,MeV}}
\def\KEV#1{10^{#1}{\rm\,keV}}
\def\lrf#1#2{ \left(\frac{#1}{#2}\right)}
\def\lrfp#1#2#3{ \left(\frac{#1}{#2} \right)^{#3}}
% UNTIL HERE

%
%\begin{flushright}
%\hfill hep-ph/yymmnnn\\
%\hfill April 2008\\
%IPMU 10-????
%\end{flushright}

\title{
Linear Inflation from Running Kinetic Term in Supergravity
}

\author{
Fuminobu Takahashi
}

\affiliation{
Institute for the Physics and Mathematics of the Universe,
University of Tokyo, Chiba 277-8583, Japan
}

\date{\today}

\begin{abstract}
  We propose a new class of inflation models in which the coefficient of
  the inflaton kinetic term rapidly changes with energy scale. This
  naturally occurs especially if the inflaton moves over a long distance
  during inflation as in the case of large-scale inflation.  The
  peculiar behavior of the kinetic term opens up a new way to
  construct an inflation model.  As a concrete example we construct a
  linear inflation model in supergravity.  It is straightforward to
  build a chaotic inflation model with a fractional power along the
  same line. Interestingly, the potential takes a different form after inflation
  because of the running kinetic term.
\end{abstract}

\pacs{98.80.Cq}

\maketitle

The inflation has been strongly motivated by the
observation~\cite{Komatsu:2010fb}, while it is a non-trivial task to
construct a successful inflation model. A successful inflaton
model must explain several features of the density perturbation, but
properties of the inflaton are not well understood. It is often
assumed that, in the slow-roll inflation paradigm, the inflaton is a
weakly coupled field, and therefore the kinetic term is simply set to
be the canonical form during inflation. This seems justified because
the typical energy scale of inflation is given by the Hubble
parameter, which remains almost constant during inflation. However,
there is another important energy scale, namely, the inflaton field.
Even in the slow-roll inflation, the motion of the inflaton is not
negligible and it may travel a long distance during the whole period
of inflation.  In particular, in the case of large-scale inflation such as a chaotic inflation
model~\cite{Linde:1983gd} , the inflaton typically moves over a Planck
scale or even larger within the last $60$
e-foldings~\cite{Lyth:1996im}. Then, it seems quite generic that the
precise form of the kinetic term changes during the course of
inflation. In some cases, the change could be so rapid, that it
significantly affects the inflaton dynamics.  In this letter, we
construct a model in which the coefficient of the kinetic term grows
rapidly with the inflaton field value, but in a controlled way. By
doing so, we construct a linear term inflation model in the
supergravity framework (see Ref.~\cite{Kawasaki:2000yn} for the
quadratic model).  The realization of the linear term inflation model
in the string theory was given in Ref.~\cite{McAllister:2008hb}. We
also show that a chaotic inflation model with a fractional power can
be straightforwardly constructed along the same line.

Before going to a realistic inflation model, let us give our basic
idea.   Suppose that the inflaton field $\phi$ has the following
kinetic term,
\beq
{\cal L}_K \;=\; \frac{1}{2} f(\phi)\, \partial^\mu \phi \partial_\mu \phi,
\label{kin}
\eeq
and that the inflaton field is canonically normalized at the potential
minimum:
\beq
f(\phi_{\rm min}) \;=\; 1.
\eeq
However, this does not necessarily mean that $f(\phi)$ remains close
to $1$ during inflation, especially if the inflaton moves over some
high scale, e.g., the GUT or Planck scale.  Suppose that the behavior
of $f(\phi)$ can be approximated by $f(\phi) \approx n^2\phi^{2n-2}$ with
an integer $n$ over a certain range of $\phi$ . Then, when expressed
in terms of the canonically normalized inflaton field, $\chi \equiv
\phi^n$, the scalar potential $V(\phi)$ is modified to be
\beq
V(\phi) \;\rightarrow \;V(\chi^{1/n}).
\eeq
For instance, if $n=2$, the quadratic potential, $V(\phi) \propto
\phi^2$, becomes a linear term $V(\chi) \propto \chi$.  Therefore,
such a strong dependence of the kinetic term on the inflaton field
changes the inflation dynamics significantly. In particular, the large
coefficient of the kinetic term is advantageous for inflation to occur,
since the effective potential becomes flatter. 

Now let us construct a linear term inflation model in supergravity. In
this inflation model, the inflaton field has a scalar potential
linearly proportional to the inflaton, $V(\phi) \propto \phi$.  For
the inflation to last for the $60$ e-foldings, the inflaton field must
take a value greater than the Planck scale, which is difficult to
implement in supergravity because of the exponential prefactor in the
scalar potential. Therefore we need to introduce some sort of  shift
symmetry, which suppresses the exponential growth of the potential.

We introduce a chiral superfield, $\phi$, and require that the
K\"ahler potential for $\phi$ is invariant under the following
transformation;
\bea
\phi^2 \;\rightarrow\;\phi^2+ \alpha~~~{\rm for~~~}\alpha \in {\bf R}
{\rm ~~~and~~} \phi \ne 0,
\label{sym}
\eea
which means that a composite field $\chi \sim \phi^2$ transforms under
a Nambu-Goldstone like shift symmetry.  
This is equivalent to imposing a hyperbolic rotation symmetry  (or equivalently SO(1,1))
on $(\phi_R, \phi_I)$, where $\phi_R$ and $\phi_I$ are the real and imaginary
components, $\phi = (\phi_R + i \phi_I)/\sqrt{2}$.

The K\"ahler potential must be
a function of $(\phi^2 - \phi^{\dag 2})$:
\beq
K\;=\; i c(\phi^2 - \phi^{\dag 2}) - \frac{1}{4} (\phi^2 - \phi^{\dag 2})^2 + \cdots,
\label{Kahler}
\eeq
where $c$ is a real parameter of $O(1)$ and the Planck unit is
adopted. Note that the $|\phi|^2$ term, which usually generates the
kinetic term for $\phi$, is forbidden by the symmetry. Instead, the
kinetic term arises from the second term, and the coefficient of the
kinetic term will be proportional to $|\phi|^2$. Note that the lowest
component of the K\"ahler potential vanishes for either $\phi_R = 0$
or $\phi_I=0$. This feature is
essential for constructing a chaotic inflation model in supergravity.

Let us add a symmetry breaking term, $\Delta K\;=\; \kappa |\phi|^2$,
to cure the singular behavior of the K\"ahler metric at the
origin. Here $\kappa \ll 1$ is a real numerical coefficient, and the
smallness is natural in the 't Hooft's sense~\cite{natural}.  
There could be other symmetry breaking terms, but,
throughout this letter we assume that those symmetry breaking terms are {\it soft}
in a sense that the shift symmetry remains a good symmetry at large enough $\phi$.
The kinetic term of the scalar field then becomes
\beq
{\cal L}_K \;=\; (\kappa + 2|\phi|^2 + \cdots) \partial^\mu \phi^\dag \partial_\mu \phi,
\eeq
where the higher-order terms expressed by the dots contain terms
proportional to $(\phi^2-\phi^{\dag2})$.  Let us drop the higher-order
terms for the moment. As demonstrated later, the higher-order terms do not
change the form of the kinetic term.  For a large field value $|\phi| \gg
\sqrt{\kappa}$, the coefficient of the kinetic term grows with the
field value, which makes the potential flatter. The canonically normalized field 
is $\chi = \phi^2/\sqrt{2}$, as expected.  In a sense, $\chi$ is a more suitable dynamical variable
to describe the system satisfying the shift symmetry (\ref{sym}).  On
the other hand, for a small field value of $|\phi| \ll \sqrt{\kappa}$,
the canonically normalized field is $\sqrt{\kappa}\phi$. Thus,
$\phi^2$ and $\phi$ are the dynamical variables for high and low
scales, respectively.

We can interpret the above phenomenon in the following way. If we go
to high energy scales, namely the large field value $\phi$, the
self-interaction in the K\"ahler potential becomes strong, and the
scalar field forms a bound state $\phi^2$. On the other hand, as
$\phi$ becomes small, the self-coupling becomes smaller and the
symmetry-breaking term becomes more relevant. Thus $\phi^2$ breaks up
and $\phi$ becomes the suitable variable.  Such a phenomenon of
forming a bound state seems quite generic if one considers a
large-scale inflation in which the inflaton takes a very large field
value during inflation. Because of the strong self-interactions, the
inflaton kinetic term runs with scales, and the inflaton dynamics is
significantly changed. The novelty here is the existence of the shift
symmetry, without which we cannot control the effect of the higher
order terms on the inflationary dynamics. We will come back to this
point later.

In order to construct a realistic inflation model, we consider the
following K\"ahler and super-potentials~\footnote{ One can add an
  additional breaking term in the superpotential which induces a
  periodic potential for $\varphi$~\cite{McAllister:2008hb}. The
  interesting feature may be found in the non-Gaussianity in this
  case~\cite{Hannestad:2009yx}.  }
\bea
K &=& \kappa |\phi|^2 +  i c(\phi^2 - \phi^{\dag 2}) - \frac{1}{4} (\phi^2 - \phi^{\dag 2})^2 + |X|^2,\\
W &=& mX\phi,
\label{W}
\eea
where both $\kappa$ and $m$ break the symmetry, and so we assume
$\kappa \ll1$ and $m\ll 1$.\footnote{The breaking of the shift symmetry in the superpotential could produce radiative corrections
to the K\"ahler potential. In particular, $\kappa = O(m^2)$ is induced~\cite{Kawasaki:2000yn}. 
Here we consider a more general case that $\kappa$ and $m$
are not related to each other.}  These small parameters are naturally
understood in 't Hooft's sense.  The superpotential produces the inflaton potential.  
We assume that $X$ and $\phi$ have U(1)$_R$ charges $2$ and $0$, respectively.
We assign a $Z_2$ symmetry under which both $X$ and $\phi$ flip the sign.

The Lagrangian is given by
\beq
{\cal L}\;=\; (\kappa + 2|\phi|^2) \partial_\mu \phi^\dag \partial^\mu \phi + \partial_\mu X^\dag \partial^\mu X -V
\eeq
with
\beq
V\;=\; e^K \left(|D_X W|^2K^{X \bar{X}} + |D_\phi W|^2 K^{\phi \bar{\phi}} - 3 |W|^2 \right).
\eeq
The scalar potential looks complicated, but it can be reduced to a simple form during inflation.
One can  show that, during inflation,  $X$ acquires a mass of the order of the Hubble scale and is stabilized
at the origin, if $|c|\gtrsim 1$.\footnote{A self-coupling $\sim |X|^4$ in the K\"ahler potential produces
a Hubble-induced mass term for $X$ about the origin.}
Then the scalar potential becomes
\beq
V\;\approx\; \frac{1}{2} e^{\frac{\kappa}{2}(\phi_R^2+\phi_I^2)-2c\phi_R\phi_I+ \phi_R^2 \phi_I^2}m^2\left(\phi_R^2+\phi_I^2\right) .
\eeq
The flat direction is given by $\phi_R \phi_I = {\rm constant}$
because of the symmetry (\ref{sym}).  Therefore if $\phi_R$ has a very
large value, $\phi_I$ is stabilized at a point where the K\"ahler
potential is minimal.  For $\phi_R > 1$ and $\kappa \ll 1$,
$\phi_I$ is stabilized at
\beq
\phi_I \;\approx \frac{c}{ \phi_R}.
\eeq
Here and in what follows we focus on the case of $\phi_R>0$ and
$\phi_I>0$ without loss of generality.  The scalar potential is then
reduced to the following form:
\beq
V\;\approx\; \frac{1}{2}e^{\frac{\kappa}{2}\left(\phi_R^2+\frac{c^2}{\phi_R^2}\right)-c^2} m^2
\left(\phi_R^2+\frac{c^2}{ \phi_R^2} \right),
\eeq
for $|\phi_R| > 1$. Since we explicitly breaks the shift symmetry
(\ref{sym}) by the $\kappa$ term, there appears a non-vanishing exponential
prefactor. However, for $|\phi_R| < 1/\sqrt{\kappa}$, the exponential prefactor is of $O(1)$,
and therefore can be dropped.\footnote{Actually, the inflaton does slow-roll if the exponential
prefactor gives a main contribution to the tilt of the potential.}
The Lagrangian for the inflaton $\phi_R$ is summarized
by
\beq
{\cal L}\;\approx\; \frac{1}{2} \phi_R^2( \partial \phi_R)^2
-  \frac{1}{2} m^2 \phi_R^2,
\eeq
for $1 \ll \phi_R \ll 1/\sqrt{\kappa}$. In terms of the canonically
normalized field, $\varphi \equiv \phi_R^2/2$, we have
\beq
{\cal L}\;\approx\; \frac{1}{2} ( \partial \varphi)^2
-   m^2  \varphi,
\eeq
for $1\ll \varphi \ll 1/\kappa$. Thus our model is equivalent to the
linear term inflation model.

\begin{figure}[t!]
\includegraphics[scale=0.5]{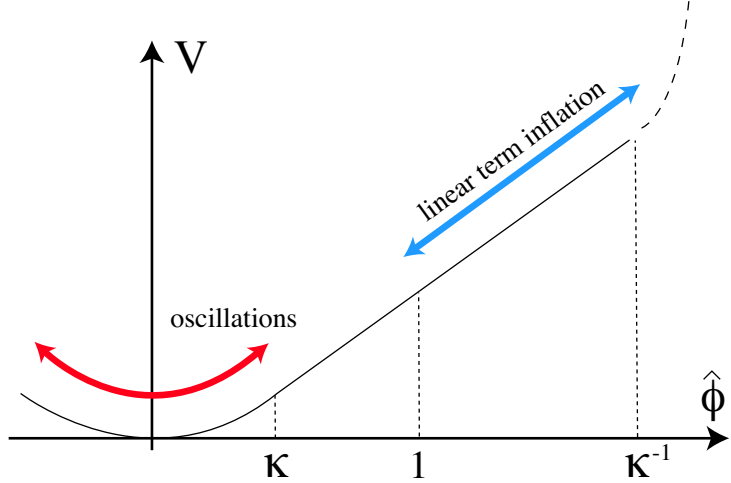}
\caption{
The schematic plot of the scalar potential in terms of a canonically normalized field, $\hat{\phi}$.
The potential is quadratic around the origin, and linear between $\hat{\phi} \sim \kappa $ and
$1/\kappa$. Note that, for $\hat{\phi} \lesssim 1$, the inflaton is a complex scalar field 
rather than a single real scalar, because the $\phi_I$ is no longer negligible.
}
\label{fig}
\end{figure}

After inflation ends, the inflaton field will oscillate about the
origin. Then the $\phi_I$ is no longer negligible, and the inflaton is
expressed by a complex scalar field.  As the amplitude decreases, the
$\kappa$-term becomes more important, and in the end, the kinetic term
arises mainly from the $\kappa$-term. The Lagrangian is
\bea
{\cal L}&\approx&  \kappa\,  \partial_\mu \phi^\dag \partial^\mu \phi 
- m^2 \phi^\dag \phi,\non\\
&=&   \partial_\mu {\hat \phi}^\dag \partial^\mu {\hat \phi }
- \frac{m^2}{\kappa} \hat{\phi}^\dag {\hat \phi},
\eea
for ${\hat \phi} \equiv \sqrt{\kappa} \phi$ much smaller than the
Planck scale. The scalar potential is schematically shown in
Fig.~\ref{fig}.
 
 %%%%%%%%%%%%%%%%%%%
%%%%%%%%%%%%%%%%%%%

The inflaton dynamics is rather simple. Using the slow-roll approximation, the inflaton field value is
parametrized by the e-folding number as 
\beq
\varphi_N \;\simeq\; 11 \sqrt{\frac{N_e}{60}},
\label{phi-N}
\eeq
and the power spectrum of the curvature perturbation is given by
\beq
\Delta_{\cal R}^2 \;\simeq\;\left(\frac{H}{\dot \varphi} \frac{H}{2 \pi} \right)^2 \simeq \frac{m^2 \varphi_N^3}{12 \pi^2}.
\eeq
Using  Eq.~(\ref{phi-N}), the mass scale $m$ is determined to be 
\beq
m \;\simeq\; 3.6\times \GEV{13} \lrfp{N_e}{60}{-3/4},
\eeq
in order to explain the WMAP result~\cite{Komatsu:2010fb}, $\Delta_{\cal R}^2 = (2.43 \pm 0.11) \times 10^{-9}$.
The physical mass of the inflaton at the potential minimum is given by
\beq
m_\phi\;=\; \frac{m}{\sqrt{\kappa}} = 4\times \GEV{14} \lrfp{\kappa}{10^{-2}}{-1/2} \lrf{m}{4 \times \GEV{13}}.
\eeq
For the linear inflation to last  for more than $60$ e-foldings, the
symmetry breaking parameter $\kappa$ must satisfy $\kappa < 0.1$. 
In the extreme case of $\kappa \sim m^2$, the inflaton mass can be as heavy as the Planck scale.

Let us discuss what happens if we include higher order terms in the
K\"ahler potential.  Suppose that the K\"ahler potential is given by
\bea
K &=& \kappa |\phi|^2 + |X|^2 + 
\sum_{n=1} \frac{c_n}{n} \,(\phi^2 - \phi^{\dag 2})^n,
\eea
where $c_n$ is a numerical coefficient of order unity.  The
inflationary path should be such that the K\"ahler potential takes a
(locally) minimum value, since otherwise the scalar potential will
blow up and no inflation occurs. There are generically multiple
inflationary paths, and, for a large enough $\phi_R$, they are given
by
\beq
\phi_I \;=\; \frac{\rm const.}{\phi_R}.
\eeq
Therefore, the kinetic term still takes a form of $|\phi|^2 |\partial
\phi|^2$, and the resultant scalar potential for a canonically
normalized field will be a linear term~\footnote{ The kinetic term
  should have a correct sign during and after inflation, which is
  realized if the K\"ahler potential is in one of the local minima
  with respect to the variation of $\phi_I$.  }. This is not
surprising, because the form of the kinetic term is determined by the
shift symmetry we imposed on a composite scalar. It is only the
symmetry-breaking terms (other than the superpotential (\ref{W}))
which make the potential deviate from the linear term.

The reheating process in supergravity inflation models have been
recently studied in a great detail~\cite{Endo:2006tf,Endo:2006qk,Endo:2007ih,Asaka:2006bv}. The
inflaton $\phi$ will get maximally mixed with the $X$ at the potential
minimum~\cite{Kawasaki:2006gs}. Therefore $\phi$ can decay into the
standard model (SM) particles through couplings of $X$ with the SM
sector. For instance, if we introduce the coupling with Higgs
doublets,
\beq
W \;=\; \lambda X H_u H_d,
\eeq
where $\lambda$ is a numerical coefficient and $H_{u(d)}$ is the
up(down)-type Higgs doublet, the reheating temperature will become
\beq
T_R \;\sim\;\GEV{10} \lrf{\lambda}{10^{-5}} \lrfp{m_\phi}{\GEV{14}}{1/2}.
\eeq
Here we have assumed that $H_u H_d$ has a $R$-charge $0$ and a negative parity 
under the $Z_2$ symmetry, and used the relation $\lambda \sim m$.
Alternatively, if we allow a symmetry-breaking term $\delta (\phi+\phi^\dag) = \delta/\sqrt{\kappa}
(\hat{\phi}+\hat{\phi}^\dag)$ in the K\"ahler potential, the inflaton
decays into the SM particles  through the gravitational
couplings with the top Yukawa interaction and the SU(3)$_C$ gauge
sector~\cite{Endo:2006qk,Endo:2007ih}. The reheating temperature will
become
\beq
T_R \;\sim\; 5 \times \GEV{6} \lrf{\delta/\sqrt{\kappa}}{10^{-3}} \lrfp{m_\phi}{\GEV{14}}{3/2}.
\eeq
Note that $\delta$ violates both the shift and $Z_2$ symmetries. 
If the inflaton mass $m_\phi$ is about $\GEV{16}$, the reheating temperature becomes about $\GEV{10}$,
which is high enough for the thermal leptogenesis to
work~\cite{Fukugita:1986hr}.  However, it is in general difficult to
satisfy the constraints from the non-thermal gravitinio problem~\cite{Kawasaki:2006gs} when
the inflaton decay is induced by the gravitationally suppressed
coupling, unless the gravitino is extremely light $m_{3/2} \leq
O(10)$\,eV~\cite{Viel:2005qj} or the gravitino is very heavy {\it and}
the R-parity is broken. We also note that the inflaton has an approximate U(1)
symmetry at the origin and may naturally acquire an inflaton number asymmetry,
which is transferred to the baryon asymmetry in the end. Also, due to the approximate
U(1) symmetry, Q-balls~\cite{Coleman:1985ki} may be formed; however, the charge is
relatively small and so it does not affect the reheating process.

It is straightforward to extend the above model to a chaotic inflation
model with a different power.  For instance, if we consider a shift
symmetry $\phi^n \rightarrow \phi^n+\alpha$ and its breaking $W = \lambda^\prime X
\phi^m$, the scalar potential for a canonical normalized field
$\varphi$ will be proportional to $\varphi^{2m/n}$ during inflation,
and to $\varphi^{2m}$ after inflation.  Such a shift symmetry on a composite field
may be realized in a non-linear sigma model. The general feature of our model is
therefore that the potential becomes steeper after inflation.
The spectral index and the
tensor-to-scalar ratio become $n_s = 1-(1+m/n)/N_e$ and $r = 8m/(n
N_e)$, where $N_e$ denotes the e-folding number.  The linear term
model corresponds to $n=2$ and $m=1$.

Due to the running kinetic term, there appears an interesting
phenomenon. For instance, consider the case of $n=m=2$. Then, it is an
usual quadratic chaotic inflation for $|\phi| > 1$, but the potential
becomes $\sim |\phi|^4$ after inflation. Therefore, the evolution of
the universe after inflation is different from the usual quadratic
chaotic inflaiton. We may take $n=m=3$, and then the inflaton
energy after inflation decreases more quickly than the radiation or
non-relativistic matter, which can lead to the enhancement of the
gravity waves or baryon asymmetry~\cite{Mukohyama:2009zs, NT}. 
This will have an important impact on the future direct gravity wave search experiments
such as advanced LIGO~\cite{Barish:1999vh}, LCGT~\cite{Kuroda:2002bg}, LISA~\cite{LISA} and DECIGO~\cite{Seto:2001qf}.
Note also that, for $m \geq 2$, the inflaton field is massless in the SUSY limit, which can relax the
thermal and non-thermal gravitino problem~\cite{NT}.

In principle, we may make use of
such a scalar field as a curvaton~\cite{curvaton} or
ungaussiton~\cite{Suyama:2008nt}.  The peculiar form of the scalar
potential may make it easier for the field to give a sizable contribution to the
total energy.

Lastly let us mention the initial condition of the inflation model.
Suppose that the $\phi$ is fluctuating in the linear potential. If the inflation lasts
for large number of e-foldings, the fluctuations will approach a
certain distribution, which should be broader than the Bunch-Davies 
distribution~\cite{Bunch:1978yq} for the quadratic potential. So the linear
inflation may take place with a larger probability.

The inflation with a running kinetic term has many implications; the potential becomes flatter,
making the inflation to occur easily, and the gravity waves can be enhanced at  frequencies within the reach
of current and future gravity wave experiments.
The future observation~\cite{:2006uk} of $n_s$, $r$, a possibly large non-Gaussianity~\cite{Hannestad:2009yx},
and direct gravity wave experiments will refute or support these models.

%%%%%%%%%%%%%%%%%%%%%%%%%%%%%%%%%%%%%%%%%%%%
\begin{acknowledgements}
%%%%%%%%%%%%%%%%%%%%%%%%%%%%%%%%%%%%%%%%%%%%
  The author thanks Martin Sloth and Christian Gross for
  discussion and Masahiro Kawasaki, Shinta Kasuya and Kazunori Nakayama for comments  
  and Antonio Riotto and CERN Theory Group for the warm
  hospitality while the present work was completed.  The work of FT
  was supported by the Grant-in-Aid for Scientific Research on
  Innovative Areas (No. 21111006) and Scientific Research (A)
  (No. 22244030), and JSPS Grant-in-Aid for Young Scientists (B)
  (No. 21740160).  This work was supported by World Premier
  International Center Initiative (WPI Program), MEXT, Japan.

 %%%%%%%%%%%%%%%%%%%%%%%%%%%%%%%%%%%%%%%%%%%%
\end{acknowledgements}
%%%%%%%%%%%%%%%%%%%%%%%%%%%%%%%%%%%%%%%%%%%%

%%%%%%%%%%%%%%%%%%%%%%%%%%%%%%%%%%%%%%%%%%%%

%%%%%%%%%%%%%%%%%%%%%%%%%%%%%%%%%%%%%%%%%%%%

\end{document}